\documentclass[aps,prb,floatfix,twocolumn]{revtex4-1}

\usepackage{amsmath,amssymb}
\usepackage[pdftex]{hyperref,graphicx}
\hypersetup{colorlinks = true, urlcolor = blue, linkcolor = blue, citecolor = blue}

\usepackage{physics}
\usepackage{xcolor}

\begin{document}
	
	\title{Generic quantized zero-bias conductance peaks in superconductor-semiconductor hybrid structures}
	
	\author{Haining Pan}
	\author{William S. Cole}
	\author{Jay D. Sau}
	\author{S. Das Sarma}
	\affiliation{Department of Physics, Condensed Matter Theory Center, Joint Quantum Institute, University of Maryland, College Park, Maryland 20742, USA}
	
	\begin{abstract}
		We show theoretically that quantized zero-bias conductance peaks should be ubiquitous in superconductor-semiconductor hybrids by employing a zero-dimensional random matrix model with continuous tuning parameters.
		We demonstrate that a normal metal-superconductor (NS) junction conductance spectra can be generically obtained in this model replicating all features seen in recent experimental results.
		The theoretical quantized conductance peaks, which explicitly do not arise from spatially isolated Majorana zero modes, are easily found by preparing a contour plot of conductance over several independent tuning parameters, mimicking the effect of Zeeman splitting and voltages on gates near the junction.
		This suggests that, even stable apparently quantized conductance peaks need not correspond to isolated Majorana modes; rather, the \textit{a priori} expectation should be that such quantized peaks generically occupy a significant fraction of the high-dimensional tuning parameter space that characterizes the NS tunneling experiments.
	\end{abstract}

\date{\rm\today}
\maketitle
\section{Introduction}
The experimental search for Majorana zero modes (MZMs) in superconductor-semiconductor (SC-SM) hybrid devices has succeeded in observing many theoretically predicted features,~\cite{nayak2008nonabelian,sarma2015majorana,alicea2012new,elliott2015colloquium,stanescu2013majorana,leijnse2012introduction,beenakker2013search,lutchyn2018majorana,aguado2017majorana,lutchyn2010majorana,oreg2010helical,sau2010generic,jiang2013nonabelian,sato2017topological,sato2016majorana,plugge2017majorana,karzig2017scalable,wilczek2012quantum,sau2010nonabelian} most notably the zero-bias conductance peak (ZBCP) and recently even the quantized ZBCP of $2e^2/h$ for normal metal-superconductor (NS) junction devices~\cite{nichele2017scaling,zhang2018quantized}.
These features are consistent with spatially isolated Majorana zero modes, but also with alternative theoretical explanations, such as ``quasi-Majorana" Andreev bound states ~\cite{kells2012nearzeroenergy,prada2012transport,liu2017andreev,vuik2019reproducing,moore2018quantized} as well as the generic localization enhancement of the density of states at zero energy in class D systems~\cite{motrunich2001griffiths,brouwer2011topological,brouwer2011probability,sau2013density,pikulin2012zerovoltage}. 
The experimental methodology for reporting candidate MZMs based on single-junction NS conductance typically involves: (1) a search over the experimental parameter space (e.g., Zeeman field, tunnel barrier, and various gate voltages to tune the chemical potential) to identify any ZBCPs with an otherwise clean, featureless spectrum below the parent superconductor gap; (2) additional parameter fine-tuning to obtain $2e^2/h$ conductance; then finally (3) a demonstration that this conductance is quantized through stability of the ZBCP as external parameters (e.g., gate voltage and Zeeman field) are tuned (i.e., a quantized conductance ``plateau").
Although some consensus already exists in the community that definitive useful information cannot be extracted at this stage from NS tunneling ZBCP measurements at a single device end~\cite{zhang2019next}, we conclusively establish that consensus by showing in this paper that even step (3) of the protocol above is generally unable to rule out the trivial non-MZM ZBCP scenario.

\begin{figure}[t]
	\centering
	\includegraphics[width=8cm]{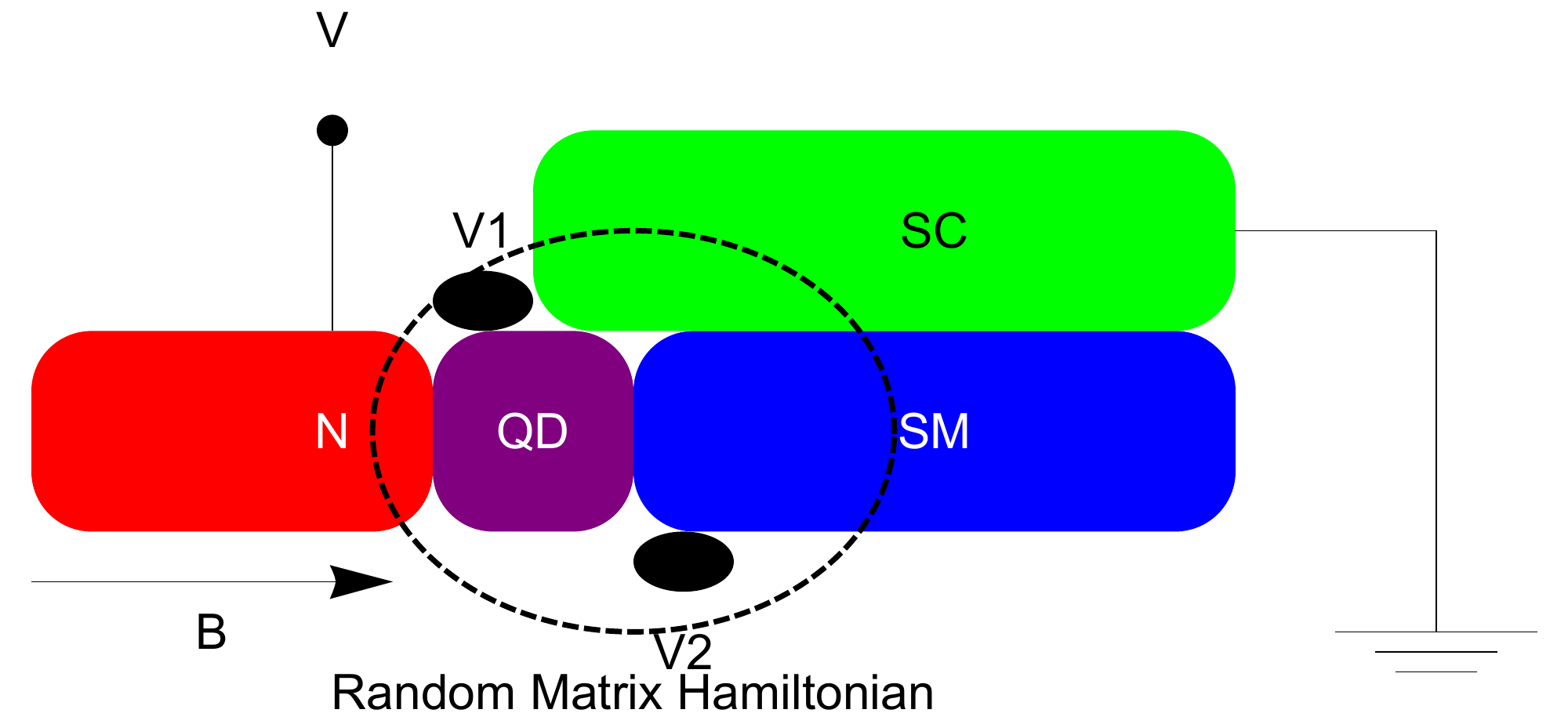}
	\caption{Schematic of a SC-SM hybrid structure coupled to a normal metal lead.  $V_1$ and $V_2$ represent two gate voltages, whereas $B$ is the applied longitudinal magnetic field. The dashed circle highlights the quantum dot (QD) region arising from the proximitized SM nanowire, which is described by the random matrix Hamiltonian.}
	\label{fig:1}
\end{figure}

Theoretically, we establish compellingly that ZBCPs of trivial origin are generic in systems with no symmetry other than particle-hole symmetry that these peaks can stick to zero energy over extended regions of the parameter space and finally that some finite fraction of these peaks manifests stable and robust quantized conductance. In other words, the experimental procedure of searching for quantized zero-bias peaks by fine-tuning experimental parameters (e.g., Zeeman field, gate voltages, and tunnel barrier) in a systematic way is practically guaranteed to produce these ``false positive" apparently quantized, but nevertheless, trivial ZBCPs.
Our theoretical starting point is a class D random matrix ensemble~\cite{altland1997nonstandard,bagrets2012class,pikulin2012zerovoltage,mi2014xshaped}. The model is maximally generic since we impose no constraint other than particle-hole symmetry on the Hamiltonian, which holds for every experimental Majorana platform. We make no claim that any previous experiment is precisely described by such a random matrix in its full details:  The random matrix theory can only predict the most generic features due to fluctuations in different Hamiltonians governed by the same symmetry (i.e., particle-hole symmetry here)~\cite{guhr1998randommatrix}. Thus, this should be only understood as a null hypothesis applied to all one-dimensional and two-dimensional Majorana platforms, including SC-SM devices~\cite{nichele2017scaling,zhang2018quantized,vaitiekenas2018effective,chen2019ubiquitous}, iron-based superconductors~\cite{wang2018evidence,zhu2019nearly}, etc.  We then calculate the NS tunneling conductance spectra of an essentially zero-dimensional quantum dot system (see Fig.~\ref{fig:1}).
Our central result expands and substantially generalizes on the generic class D ``sticking ZBCP"~\cite{mi2014xshaped} scenario--- i.e., a nonquantized conductance peak that remains at zero bias as a single continuous parameter is tuned (this was also called a ``\textit{Y}-shaped Andreev resonance" in Ref.~\onlinecite{mi2014xshaped}). In a higher-dimensional parameter space as consistent with the experimental methodology searching for MZMs, we show here that sticking ZBCPs can evolve into extended plateaulike regions enclosed by contours with $2e^2/h$ conductance. We establish that it is typical to find completely generic random Hamiltonians that produce sticking ZBCPs over some fraction of the parameter space \emph{and} with stable near-quantized conductance but which never correspond to spatially isolated MZMs by construction. If we are allowed some additional postselection over the space of random class D matrices as can be performed by tuning several parameters in the experimental protocol, these plateaus can even be made remarkably large. We also show that varying the tunneling amplitude leads to qualitatively similar results to those reported in experiments by changing the barrier gate voltage~\cite{zhang2018quantized}.

We choose the Gaussian ensemble to simulate the quantum dot so that each element of the random Hamiltonian is distributed independently, which simplifies the later calculation~\cite{altland1997nonstandard,beenakker1997randommatrix}. We mention also that our random matrix model, whereas being the most generic theoretical model of the hybrid system from the perspective of symmetry, is also a reasonably physical model of the currently available experimental samples where the nanowire is typically short with many discrete energy levels being occupied. This guarantees the prerequisite of a valid random matrix approach that the energy states in the Hamiltonian spectrum should be sufficiently large~\cite{beenakker1997randommatrix}. In essence, nanowires that have a high chemical potential and, thus, occupy many subbands act like an effective random matrix system because any changes in the system parameters (e.g., various gate voltages) can drastically alter the Hamiltonian of the system during the measurement, even though the material itself is in the clean limit~\cite{woods2019subband}.

\section{Model and formalism}
The region in Fig.~\ref{fig:1} enclosed by the dashed line is mainly composed of a zero-dimensional superconducting quantum dot appearing in a proximitized SM nanowire, which can be described by a class D $M \times M$ Hermitian random matrix. In the following, the Majorana basis is adopted for simplicity. In this basis, the class D ensemble is characterized by a particle-hole symmetry,
\begin{equation}\label{eq:symm}
H=-H^*.
\end{equation}
It is then convenient to take $H=iA$, where $A$ is real and anti-symmetric. In the large-$M$ limit~\cite{mi2014xshaped,beenakker1997randommatrix,dittes2000decay,guhr1998randommatrix,altland1997nonstandard}, we can assume a Gaussian distribution for $H$,
\begin{equation}\label{eq:gaussian}
P(H) \propto  \exp(-\frac{c}{M}\tr(H^2)), 
\end{equation}
where $\tr(\cdots)$ is the matrix trace and $c=\pi^2/(4\delta_0^2)$.
The parameter $\delta_0$ is the mean energy-level spacing.
Equation~\eqref{eq:gaussian} can be further simplified to independent Gaussian distributions for each element of $A$,
\begin{equation}\label{eq:P(A)}
P(\left\{A_{nm}\right\})\propto \prod_{1=n<m}^{M} \exp\left(-\frac{\pi^2 A_{nm}^2}{2M \delta_0^2}\right).
\end{equation}

\begin{figure}[t]
	\centering
	\includegraphics[width=8cm]{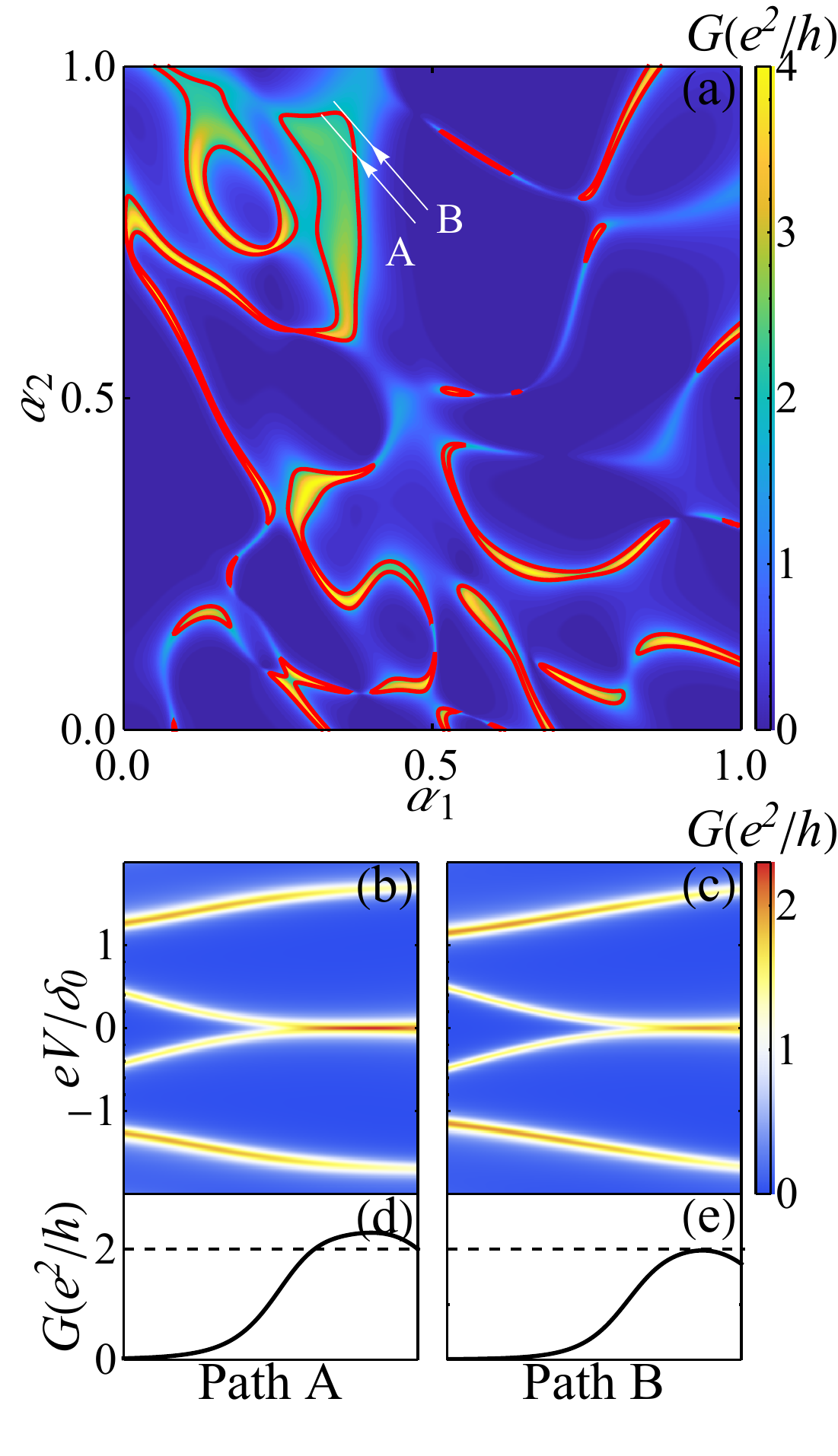}
	\caption{
	(a) Color plot of the differential conductance at zero-bias voltage $G(0)$, as a function of tuning parameters $\alpha_1$ and $\alpha_2$. The red contour line joins all points with conductance $2e^2/h$. We set the temperature $T=0$.
	The marked lines $A$ and $B$ indicate paths of varying $\alpha$ mimicking the tuning of gates in experiments.
	(b) The conductance as a function of the bias voltage and position along path $A$. The bias voltage is normalized by the mean level spacing $\delta_0$. This is an example of the gap closing feature followed by trivial nearly quantized ZBCP.
	(c) The same along path $B$. (d) and (e) are the corresponding ZBCP heights.} 
	\label{fig:2}
\end{figure}

To calculate the conductance, we follow the formalism of Ref.~\onlinecite{mi2014xshaped}, which we reproduce here to make the presentation self-contained. To simulate the NS tunneling geometry, the $M$ states of the quantum dot are coupled to a $N$-channel lead through the $M \times N$ tunneling matrix $W$. Since the choice of the basis of $H$ does not affect the distribution of $H$, we can, without loss of generality, simply choose the basis where the tunneling matrix $W$ is  diagonal~\cite{mi2014xshaped}, i.e.,
\begin{equation}\label{eq:W}
W_{mn}=w_n\delta_{m,n}, \quad 1\le m\le M, \quad 1\le n \le N,
\end{equation}
where $ w_n $ is determined by the tunneling probability $\Gamma_n \in [0,1]$ as~\cite{guhr1998randommatrix,beenakker1997randommatrix}
\begin{equation}\label{eq:wn}
\abs{w_n}^2=\frac{M\delta_0}{\pi^2\Gamma_n}\left(2-\Gamma_n-2\sqrt{1-\Gamma_n}\right).
\end{equation}
For simplicity, we assign an identical tunneling probability for each channel in the lead, following Ref.~\onlinecite{mi2014xshaped}.

The differential conductance $G(V)=dI/dV$ is then determined by calculating the $N \times N$ scattering matrix~\cite{setiawan2015conductance,prada2012transport,liu2017andreev}
 according to the Mahaux-Weidenm\"uller formula~\cite{mahaux1969shellmodel,christiansen2009mathematical,marciani2014timedelay,guhr1998randommatrix,beenakker1997randommatrix},
\begin{equation}\label{eq:S}
S(E)=1+2\pi W^\dagger(H-i \pi W W^\dagger-E)^{-1} W.
\end{equation} 
We then obtain the differential conductance in the Majorana basis as
\begin{equation}\label{eq:G}
G(V)=\frac{e^2}{h}\left(\frac{N}{2}-\frac{1}{2}\tr[S(eV)\tau_yS(eV)^\dagger\tau_y]\right)
\end{equation}
where $\tau_y$ is the Pauli matrix acting on the particle-hole space.

Another useful tool for understanding the sticking of ZBCPs is the non-Hermitian ``effective Hamiltonian", 
\begin{equation}\label{eq:Heff}
H_{\text{eff}}=H-i\pi W W^\dagger,
\end{equation}
where the imaginary term is a self-energy acquired from the coupling to the lead. When the tunneling probability $\Gamma_n$ is small, the energy spectrum of $H_{\text{eff}}$ approaches that of the original $H$. Note that the eigenvalues of $H_{\text{eff}}$ are distributed in the lower half of the complex plane due to the positive definiteness of $W W^\dagger$. In addition, the particle-hole symmetry in $H$ constrains the eigenvalues to come in pairs $\epsilon$ and $-\epsilon^*$. Therefore, the eigenvalues will be symmetrically distributed along the imaginary axis, unless purely imaginary. Thus, the nondegenerate eigenvalue on the imaginary axis has a range of stabilities \cite{pikulin2013two,pikulin2012topological} against perturbations since there is no way to obtain a nonzero real part without breaking the particle-hole symmetry. This kind of stability is responsible for the zero energy sticking of the trivial ZBCPs. An intuitive way to understand this is that a purely imaginary eigenvalue of $H_{\text{eff}}$ corresponds to the presence of an exact zero eigenvalue in $H$. Consequently, one will observe a corresponding ZBCP, especially when $\Gamma_n$ is small.
In what follows, we use $M=80$ and $N=4$ (i.e., a single transport channel with particle-hole and spin degrees of freedom). Our results are sensitive to the choice of $N$, so we choose it to reflect the experimental situation; our results are not sensitive to the choice of $M$, once sufficiently large. The tunneling probability is $\Gamma_n = 0.1$ by default unless stated otherwise. We emphasize that all of our results, quantized conductance or not, are, by construction, topologically trivial since $M$ is even~\cite{ryu2010topological,hasan2010colloquium,qi2011topological}.

\begin{figure}[t]
	\centering
	\includegraphics[width=8cm]{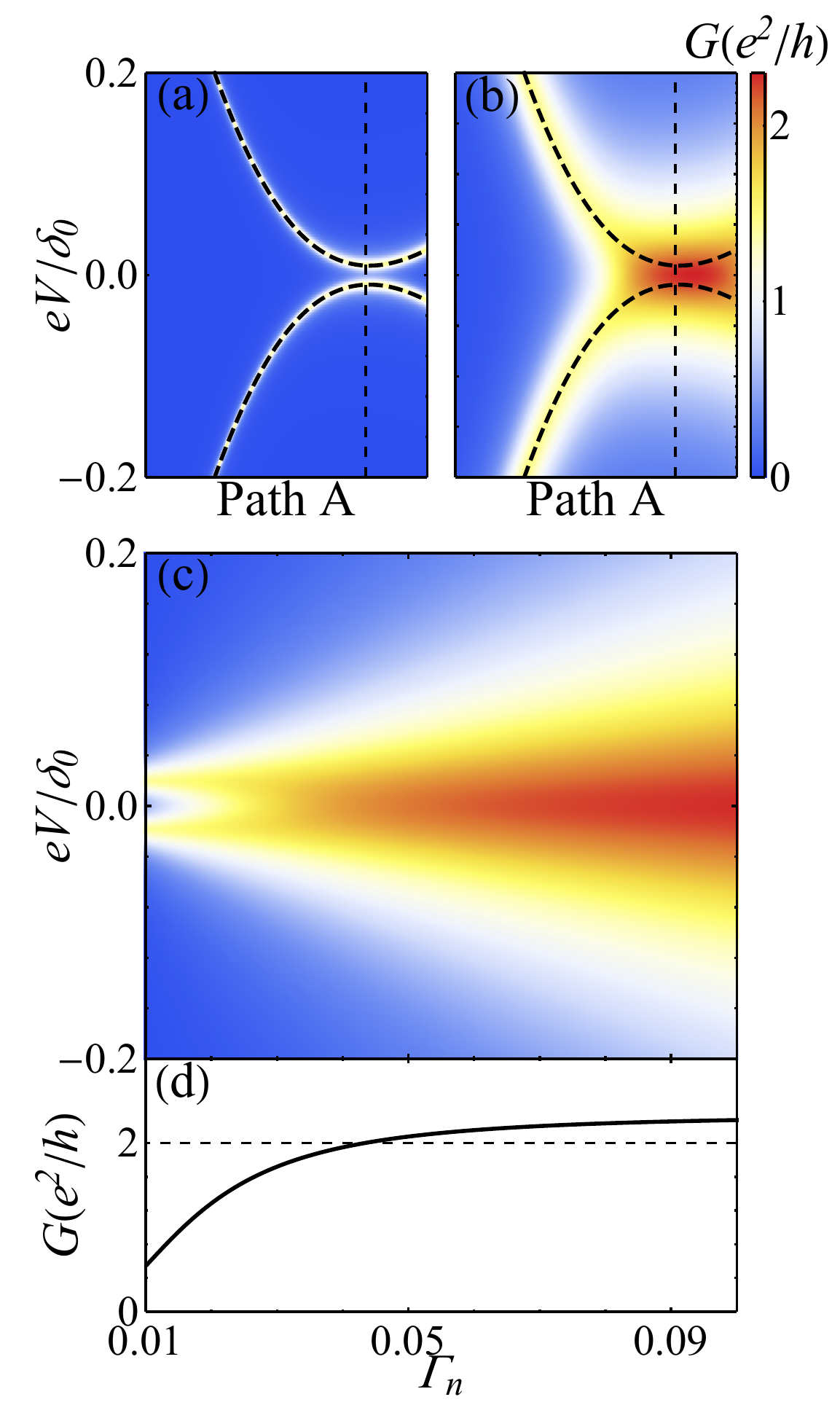}
	\caption{Conductance spectra near zero energy along path $A$ as in Fig.~\ref{fig:2} for (a) $\Gamma_n=0.01$ and (b) $\Gamma_n=0.1$. The dashed lines are the eigenvalues of $H$. For sufficiently large $\Gamma_n$, the corresponding conductance smears the level structure into a stuck ZBCP. Fixing $\alpha_{1,2}$, the conductance as a function of transparency is shown in (c), and the ZBCP height is shown in (d).}
	\label{fig:3}
\end{figure}

Finally, our conceptual break with Ref.~\onlinecite{mi2014xshaped} is that we simulate the manipulation of several independent parameters (simulating the role of gate voltages in experiments), describing the quantum dot with a parametric Hamiltonian in a higher-dimensional parameter space (i.e., Zeeman field plus gate voltages),
\begin{equation}\label{eq:H}
H=\alpha_1 H_1+ \alpha_2 H_2+ (1-\alpha_1-\alpha_2) H_3,
\end{equation}
where $H_{1-3}$ are three randomly drawn matrices and $\alpha_{1,2}\in[0,1]$ are two tuning parameters. This two-dimensional parameter space can be considered as an isomorphism to a space spanned by magnetic-field $B$ and another gate voltage ($V_{1}$ or $ V_2 $) in the experiment through a certain deterministic mapping (which does not have to be orthogonal or linear). However, we also clarify that, since there are typically several gates in experiments, the manner in which we introduce two parameters $\alpha_{1,2}$ does not forbid the possibility of a higher-dimensional parameter space. In fact, it is very straightforward to generalize the two-parameter-dependent Hamiltonian to a higher-dimensional one. We find, however, that the two-dimensional parameter space is already sufficient to establish our main conclusions, and any higher-dimensional parameter space representation only reinforces the generic results presented here.

\section{Results}
We now emulate the ``Majorana search" protocol of tuning parameters to find sticking ZBCPs. We do this systematically by finding the purely imaginary eigenvalues of Eq.~\eqref{eq:Heff}. We randomly draw 125000 independent triplets for $H_{1-3}$, and then we post-select a $ H $ that maximizes the region of $\alpha_{1,2}$ with sticking ZBCP by choosing the particular set of $H_{1-3}$ which yields the largest number of purely imaginary eigenvalues. To determine the occurrence frequency of purely imaginary eigenvalues over realizations of $H$, we discretize the parameter space $\alpha_{1,2}\in[0,1][0,1]$ and sweep over each grid point of $\alpha_{1,2}$ to evaluate the corresponding $H$ in Eq.~\eqref{eq:H}. By enumerating the occurrences of purely imaginary eigenvalues, we indirectly know the likelihood that one can find the sticking ZBCP in such a configuration. In Fig.~\ref{fig:2}(a), we plot the differential conductance at zero-bias voltage $G(0)$ for the $H$ selected by this procedure. The conductance varies continuously from 0 (dark blue) to $4e^2/h$ (light yellow), the maximum possible for $N=4$~\cite{sengupta2001midgap}. Thus, searching for the sticking quantized ZBCP is equivalent to finding the region of $2e^2/h$ conductance in the color plot. To compare with experimental results~\cite{zhang2018quantized,vaitiekenas2018effective,chen2019ubiquitous,nichele2017scaling}, we highlight the contour of $2e^2/h$ conductance in Fig.~\ref{fig:2}(a) in red. From this, we can immediately identify two types of regions with high conductance: the \emph{plateaulike} regions on the upper left versus the \emph{ridgelike} regions on the right of Fig.~\ref{fig:2}(a). The plateaulike regions are characterized by small gradients in the conductance and are reminiscent of the theoretical expectation for a topological region of a phase diagram. In Figs.~\ref{fig:2}(b) and 2(c), we show the full conductance spectra $G(V)$ along the two paths marked $A$ and $B$ in the plateaulike region of Fig.~\ref{fig:2}(a). The conductance spectra in Figs.~\ref{fig:2}(b) and (c) show a remarkable resemblance to experimental candidate Majorana ZBCPs~\cite{zhang2018quantized,chen2019ubiquitous,vaitiekenas2018effective,nichele2017scaling}, i.e., an in-gap conductance peak coming down to zero-bias voltage, followed by a stable ZBCP of $2e^2/h$. More importantly, this nearly quantized ZBCP can even be stable in a certain region of parameters. Shifting path $A$ slightly to obtain path $B$, the conductance spectrum is hardly affected as long as the path remains within the plateaulike region. These ZBCPs in the plateaulike region can manifest a ``robust nature" in experiments. Again, we emphasize that, by construction, these nearly quantized ZBCPs shown in the conductance spectra do not arise from a spatially isolated MZM. Furthermore, these trivial nearly quantized ZBCPs are ubiquitous in Fig.~\ref{fig:2}; plateaulike regions are not rare in the color plot, and any path that crosses a plateaulike region (such as paths $A$ and $B$) will result in similar conductance spectra to Figs.~\ref{fig:2}(b) and 2(c).

We also investigate the effect of tuning the barrier transparency through $\Gamma_n$. We first vary $\Gamma_n$ from $0.01$ to $0.1$ along path $A$ of Fig.~\ref{fig:2}(a). In Figs.~\ref{fig:3}(a) and 3(b), we compare the conductance spectra for two values of $\Gamma_n$, overlaid with the eigenvalues of $H$. As $\Gamma_n$ increases, the ZBCP broadens and appears more stuck to the zero voltage axis~\cite{mi2014xshaped},  whereas for small $\Gamma_n$, the peak becomes steeper and fainter, and the origin of this ZBCP as a parabolic near touching becomes apparent. This phenomenon is consistent with experiment~\cite{zhang2018quantized} in the sense that the zero-bias conductance looks nearly quantized for larger transparency but vanishes at small $\Gamma_n$ where it is accompanied by a peak splitting as shown in Fig.~\ref{fig:3}(c).

\begin{figure}[t]
	\centering
	\includegraphics[width=8cm]{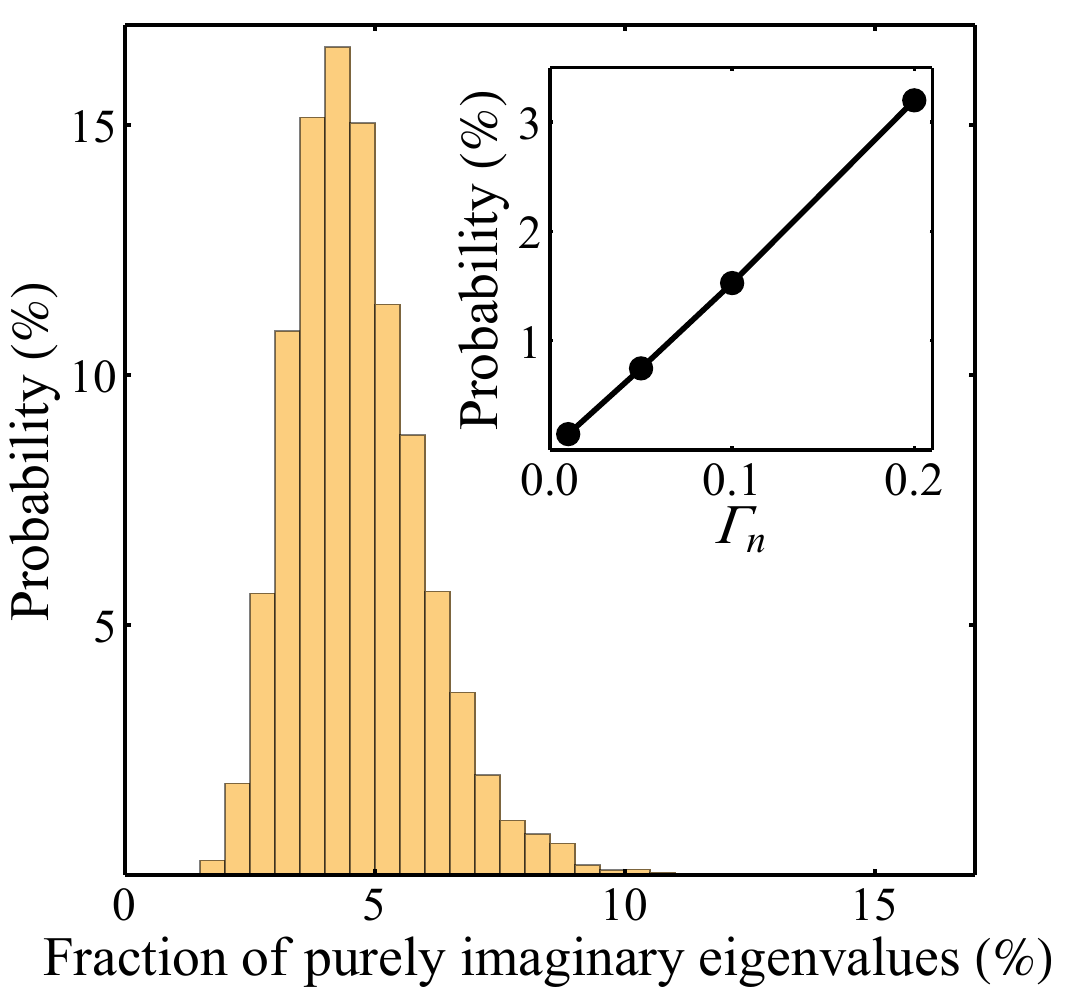}
	\caption{Distribution (over the ensemble of $H$) of the fraction of the parameter space $(\alpha_1, \alpha_2)$ where $H_{\text{eff}}$ [Eq.~\eqref{eq:Heff}] possesses purely imaginary eigenvalues. This is proportional to the area of parameter space covered by plateaulike ZBCPs. The inset shows the almost-quantized ZBCP statistics as a function of $ \Gamma_n $ which is determined by the tunnel barrier voltage.} 
	\label{fig:4}
\end{figure}

Although the previous plots were all produced for a single particular realization of $H$, we emphasize that the qualitative behavior is typical as we have verified explicitly: Our main results are that the existence of plateaus is generic and these plateaus can be easily obtained by following the generally accepted Majorana search experimental methodology. To quantify this, we calculate the fraction of the parameter space $(\alpha_1, \alpha_2)$ where $H_{\text{eff}}$ has purely imaginary eigenvalues, which, in turn, roughly corresponds to the fraction of parameter space covered by plateaulike ZBCPs. We histogram this fraction over independent realizations of $H$, shown in Fig.~\ref{fig:4}. This coverage fraction has a distribution that is peaked around 4\% --- i.e., when performing a search for MZMs through a two-parameter conductance map (for example, to construct a topological phase diagram), one should \textit{a priori} expect around 4\% of the map to feature trivially almost-quantized conductance. In the inset of Fig. \ref{fig:4}, we show the typical statistics for the expected values of these ZBCPs within $ \pm 20\% $  of the putative quantized value of $ 2e^2/h $.  Clearly, apparently quantized stable trivial ZBCPs are generic in class D superconducting systems.

\section{Discussion and conclusion}
We have generalized the crucial insight of Ref.~\onlinecite{mi2014xshaped} that distinguishes \textit{X}- versus \textit{Y}-shaped Andreev resonances (the latter being our sticking ZBCPs) to a higher-dimensional parameter space where the distinction becomes one between ridgelike and plateaulike regions in the parameter space. Although \textit{Y}-shaped resonances of a single parameter might not be quantized in general, the quantized plateaulike higher-dimensional regions are generic, and if we allow even more parameters $\alpha_i$, the parameter space that realizes nearly quantized trivial ZBCPs can only increase. This implies that more gate voltages being independently tuned to optimize experimental ZBCPs necessarily leads to the observation of generic and trivial quantized ZBCPs! Our simulated results for conductance spectra in the zero-dimensional random matrix model resemble the existing experimental NS conductance results, regardless of the proposed Majorana or quasi-Majorana interpretations. The nearly quantized ZBCP is, therefore, ubiquitous in theory in multidimensional parameter space, requiring no input other than class D symmetry--- a symmetry that any platform pursuing Majorana should possess--- and can be easily observed in conductance color plots, such as Fig.~\ref{fig:2}(a), i.e., modeling the multiparameter tuning involved in Majorana search procedures.

Before concluding, we comment on the applicability of our random matrix theory based considerations to realistic Majorana nanowire systems, where the best existing devices~\cite{nichele2017scaling,zhang2018quantized} use semiconductor nanowires of high quality, which should be reasonably free of disorder.  It may be worthwhile to emphasize that even the best InSb or InAs nanowires would typically have a mobility of 50000 $ \text{cm}^2 \text{V}^{-1}\text{s}^{-1} $ which would correspond to around one charged impurity for every 20-nm length of the wire, leading to, at least, 25 charged impurities (impurity density $ \sim 10^{16} \text{cm}^{-3} $) per nanowire currently used in Majorana experiments.  The randomness in the spatial location and the potential strength of these charged impurities would, in principle, lead to a random matrix quantum dot-type behavior even in the isolated nanowire at low temperatures with many localized random energy levels in the wire.  The actual experimental situation may be worse since the nanowire in the semiconductor-superconductor hybrid system is likely to be more disordered because of the presence of the superconducting metal and various gates and contacts.  What is even more noteworthy is the point mentioned in the Introduction that the Majorana nanowire system may act like a random system even without any strong disorder simply by virtue of many occupied subbands in the system under experimental conditions~\cite{woods2019subband}.  It seems that current Majorana nanowires in the hybrid system may have $ \sim $20--50 occupied subbands, which would lead to a natural random matrix description as was pointed out for nuclear energy spectra (and, in general, for the statistical energy level distribution in all complex systems) a long time ago~\cite{wigner1955characteristic,dyson1962statistical}.   The multisubband occupancy of Majorana nanowires makes a random matrix description of the type used in our theory a natural theory for the existing samples even if the actual disorder is not strong, although, in practice, the actual disorder in the currently available samples is likely to be large also.

Our results clarify that even stable apparently quantized ZBCPs do not provide conclusive evidence for topological MZMs; the observation of such peaks does not necessarily imply a topological phase nor does the peak merging or splitting as a function of electrostatic gate manipulation (especially local gates near the junction) imply entering or exiting such a phase. We emphasize that our theory applies to all class D superconducting platforms where ZBCPs have been reported, which include, in addition to the well-studied semiconductor nanowires~\cite{nichele2017scaling,zhang2018quantized,zhang2019next}, also ferromagnetic chains~\cite{dumitrescu2015majorana}, topological insulator-based structures~\cite{xu2014artificial}, and Fe-based superconductors~\cite{wang2018evidence,zhu2019nearly}.

We thank S. Frolov and H. Zhang for helpful discussions. This work was supported by the Laboratory for Physical Sciences and Microsoft. We also acknowledge support of the University of Maryland supercomputing resources~\cite{hpcc}.

\bibliography{RandomMatrix}
\end{document}